\documentstyle[12pt]{report}

\thicklines                         
\let\br=\u                      

\def\a{\alpha}
\def\b{\beta}
\def\c{\chi}
\def\d{\delta}
\def\e{\epsilon}                
\def\f{\phi}                    
\def\g{\gamma}
\def\h{\eta}

\def\j{\psi}
\def\k{\kappa}
\def\l{\lambda}
\def\m{\mu}
\def\n{\nu}
\def\o{\omega}
\def\th{\theta}                  
\def\s{\sigma}                  
\def\t{\tau}
\def\u{\upsilon}
\def\x{\xi}
\def\z{\zeta}
\def\D{\Delta}
\def\F{\Phi}
\def\G{\Gamma}

\def\L{\Lambda}

\def\S{\Sigma}

\def\cd{{\cal D}}

\def\cf{{\cal F}}

\def\bo{\raisebox{-.4ex}{\large$\Box$}}                 
\def\cbo{{\,\raise-.15ex\Sc [\,}}                       

\def\pa#1{\partial_{#1}}                        
\def\sl#1{\rlap{\hbox{$\mskip 1 mu /$}}#1}      
\def\Tilde#1{\widetilde{#1}}                    
\def\Hat#1{\widehat{#1}}                        
\def\Bar#1{\overline{#1}}                       
\def\dt#1{{\buildrel {\hbox{\large .}} \over {#1}}}     
\def\ddt#1{{\buildrel {\hbox{\LARGE .\kern-2pt.}} \over {#1}}}

\def\beqn#1{ \renewcommand{\theequation}{#1}
             \begin{equation} }
\def\eeqn{\end{equation}}
\def\beqr#1{ \renewcommand{\theequation}{#1}
             \begin{eqnarray} }
\def\eeqr{\end{eqnarray}}
\def\NON{\nonumber\\}
\def\beqrabc#1{ \setcounter{equation}{0}
                \renewcommand{\theequation}{#1\alph{equation}}
                \begin{eqnarray} }

\def\Int#1{\int\! d#1\,}

\def\frac#1#2{ {\sstyle {#1\over #2} } }

\def\NPB#1{Nucl. Phys. {\bf B#1}}
\def\PLB#1{Phys. Lett. {\bf B#1}}
\def\PRD#1{Phys. Rev. {\bf D#1}}

\def\PRP#1{Phys. Rep. {\bf #1}}

\def\sstyle{\scriptstyle}

\def\rhs{\mbox{r.h.s.} }
\def\lhs{\mbox{l.h.s.} }
\def\ie{\mbox{i.e.} }
\def\eg{\mbox{e.g.} }

\def\half{{1\over 2}}

\begin{document}
\noindent July 1992 \hfill IC/92/131
\par
\begin{center}
\vspace{15mm}
{\large\bf Compensating fields and \\ Anomalies in Supergravity}\\[5mm]
{\it by}\\[5mm]
Yigal Shamir$\,$\footnote[1]{Address after August 1st:
Department of Physics, Weizmann Institute of Science, Rehovot 76100, ISRAEL.}\\
{\it International Center for Theoretical Physics}\\
P.O.Box 586, 34100 Trieste, ITALY\\[15mm]
{ABSTRACT}\\[2mm]
  \end{center}
\begin{quotation}
  We discuss the quantization of theories which are formulated using
compensating fields. In particular, we discuss the relation between the
components formulation and the superspace formulation of supergravity theories.
The requirement that the compensating field
can be eliminated at the quantum level gives rise to on-shell constraints on
the operators of the theory. In some cases, the constraints turn out to be
physically unacceptable. Using these considerations we show that new minimal
supergravity is in general anomalous.
\end{quotation}

\newpage
\noindent {\bf 1.~~Introduction}
\vspace{3ex}

  In this paper we discuss the quantization of theories
which are formulated using compensating fields. We focus in particular on
the relation between the components formulation and the superspace formulation
of supergravity theories. We show that new minimal supergravity
is in general anomalous and discuss the relevance of our results to old
minimal supergravity.

  We will denote as compensating field (or compensator for short)
any field which can be eliminated
{\it algebraically} by fixing the invariance under some local symmetry.
We then refer to the local symmetry as unphysical.
(See refs.~[1,2] for a general discussion on compensators).
Notice that the algebraic symmetry structure completely determines
which fields are compensators.
In the following, it will also be useful to distinguish between propagating and
non-propagating compensators. The former have kinetic energy terms in the
lagrangian
whereas the latter do not. Thus, the distinction between the two types of
compensators depends on the {\it dynamics}.

  As we will show below, to every compensator there corresponds a constraint or
a consistency requirement at the quantum level. The constraint
requires that the inhomogeneous term in the field equation of
the compensating field vanish on the physical Hilbert space. (The
constraint is empty if the action does not contain any term linear in the
compensating field). As we will see, the constraint expresses the
quantum invariance under the unphysical symmetry needed to eliminate the
compensator. It arises because the peculiar transformation law of the
compensator allows us to identify it with the parameter of the unphysical
local symmetry.

  In this paper we discuss in detail compensators for U(1)
chiral symmetry.  For a propagating compensator we find that
the constraint is physically sensible. It expresses the longitudinal
component of a massive axial vector field in terms of other fields.
But for a non-propagating compensator the constraint
requires that the anomaly {\it operator} be zero which is
physically unacceptable.

  This example may reflect a more general pattern, namely, constraints
generated by propagating compensators are more likely to be physically
sensible.
In an anomalous theory, the introduction of a non-propagating
compensator allows us to trade one inconsistency with another.
Without the compensator we face quantum breakdown of the invariance under a
local symmetry. With the compensator we can restore the invariance under the
local symmetry by a counter-term, but we end up with a constraint which gives
rise to a trivial Hilbert space when enforced on the operators of the theory.
We stress, however, that at the moment we have no general
proof that all constraints generated by propagating compensators are
physically acceptable. Thus, one must examine the constraints generated
by all compensators of a given theory on a case by case basis.

  The freedom to reformulate a theory using compensators limits
considerably the usefulness of the cohomological approach to anomalies.
Anomalies in field theory are associated
with the existence of non-trivial cocycles of an appropriate BRST operator.
The cohomological program treats all symmetries and fields on the
same basis. Consequently, the introduction of a compensating field changes the
cohomological problem. For instance, if one introduces a conformal
compensator, the trace anomaly becomes cohomologically
trivial~[2]. The question arises where has the information about the anomaly
gone.
Our answer is that it is now contained in the on-shell constraint that
follows from the invariance under dilatations which is now maintained also
at the quantum level (though at the price that dilatation is an
unphysical symmetry). We anticipate that the constraint will be physically
acceptable if and only if the theory is non-anomalous in the usual sense when
formulated
without the compensator. In the case of a conformal compensator, we expect that
the constraint will be physically acceptable for Einstein gravity but
physically unacceptable for Weyl gravity.

  Our general discussion on compensators is particularly relevant to
supersymmetric (SUSY) theories which have some local gauge invariance.
SUSY theories can be formulated in several different ways.
The most economic formulation contains only physical fields.
Its disadvantage is that the SUSY algebra closes only on-shell.
In order to obtain a closed SUSY algebra off-shell one introduces
auxiliary fields. We refer to this formulation as the ``components
formulation''. Its characteristic property is that the anticommutator of two
SUSY transformations contains not only a translation but also gauge
transformations with field dependent coefficients. Thus, one has structure
functions instead of structure constants.

  A third formulation of SUSY theories is the superspace formulation.
For theories with some local gauge invariance
the superspace symmetry group is larger and, at the same time,
has a simpler structure
than the symmetry group of the components formulation.
The enlargement of the local symmetry group
is achieved at the cost of introducing a large number of compensating fields.
In order to go back to the components formulation one
eliminates the compensating fields {\it algebraically} by fixing the invariance
under the extra local symmetries of the superspace formulation.
This process is usually called going to the Wess-Zumino gauge.

  At the classical level, all three formulation are clearly equivalent.
We expect that the same statement should remain true at the quantum level.
But in order to establish the quantum equivalence
one must first study what are the precise
conditions needed for quantum consistency in each of these
formulations. As follows from our introductory remarks, {\it invariance under
all local symmetries is not a sufficient condition for quantization
in the presence of compensating fields}. In addition, one must verify that
the on-shell constraints, which follow from the invariance under the
unphysical symmetries, can be imposed on the operators of the theory without
leading to a trivial Hilbert space.

  This paper is organized as follows. In Section~2 we
consider a simple example based on the chiral anomaly.
The introduction of a compensator allows us to cancel the anomaly with a
Wess-Zumino term (at the cost of making the local chiral symmetry unphysical).
But for a non-propagating compensator the resulting theory is still
inconsistent because the Wess-Zumino term gives rise to a physically
unacceptable constraint.

  In section~3 we show that the invariance under an
unphysical symmetry gives rise to an on-shell constraint. A detailed proof
is given for the case of an abelian unphysical symmetry.
We expect, however, that analogous results should hold in general.

  In section~4 we discuss {\it new minimal} supergravity~[3] in its component
formulation. What characterizes new minimal supergravity is
the presence of local U(1) chiral symmetry which is gauged by an auxiliary
field. We first discuss
the differences between new minimal and old minimal supergravity.
We then show that local SUSY is anomalous in new minimal supergravity whenever
the chiral symmetry is anomalous. If the axial R-current is
not conserved the SUSY current is not conserved too.

  In section~5 we discuss the relation between the components formulation and
superspace formulation. Gates, Grisaru and Siegel~[1,4] have studied in detail
the various supergravity models in superspace. Our results concerning new
minimal supergravity agree with their conclusions, and we do not repeat their
analysis here.

  We show that invariance under the extra local symmetries of the
superspace formulation implies the {\it equality} of the superspace and
components' generating functionals (except for an unimportant normalization
constant). Thus, it is impossible that the superspace generating functional
will describe a consistent theory if the components formulation
generating functional does not. However, in superspace the inconsistency
may either take the form of quantum non-invariance under a local symmetry,
or be a consequence of an
on-shell constraint which cannot be enforced on the operators of the theory.
We conclude with some comments on the
issue of the consistency of {\it old minimal} supergravity~[5].

\newpage
\noindent {\bf 2.~~An example}
\vspace{3ex}

  We begin with a simple example which illustrates
how careless treatment of
a model containing a compensating field can lead to inconsistencies.

  Consider a single free massless Weyl fermion, whose action
$$S_0= {i \over 2} \Int{^4x}\Bar{\j}(1-\g_5)\sl{\partial} \j \,, \eqno(2.1)$$
is classically invariant under global U(1) rotations. As is well known,
requiring Lorentz invariance and Bose symmetry one finds that
current conservation is violated in the correlator
of three left-handed currents.

  We now  minimally couple the fermion to an external gauge field
$$ S_{\j} = S_0 - g \Int{^4x} A^{\m}(x) j_{\m}(x) \,,\eqno(2.2)$$
$$ j_{\m} = \half\,\Bar{\j}(1-\g_5)\g_{\m}\j \,,\eqno(2.3)$$
thus promoting the global symmetry to a local one at the classical level.
An infinitesimal gauge transformation of the effective action
$$ \G(A_{\m}) = -i\log Z(A_{\m}) \,,\eqno(2.4)$$
$$ Z(A_{\m}) = \int \cd\Bar{\j} \cd\j\, \exp \{iS_{\j}\}
\,, \eqno(2.5)$$
is induced by the BRST operator
$$ \S=\Int{^4x} \d_{\s}A_{\m}(x)\, {\d\over\d A_{\m}(x)} \,,\eqno(2.6)$$
$$ \d_{\s}A_{\m}(x) = {1\over g} \pa{\m}\s(x) \,,\eqno(2.7)$$
where $\s(x)$ is a dimensionless anticommuting ghost field. The
inability to maintain current conservation at the quantum level
manifests itself through the lack of gauge invariance of the effective action
$$ \S\,\G \equiv \Int{^4x}\s(x)\D(x) \ne 0 \,. \eqno(2.8)$$
In this example of an abelian symmetry, the requirement that the anomaly
$\D(x)$ be a non-trivial cocycle means that $\D(x)$ should be invariant
under local chiral transformations.
The only candidate with the right dimension and parity is
$F\Tilde{F}(x)$. Thus,
$$ \D(x) = c g^2  F\Tilde{F}(x) \,,\eqno(2.9)$$
where $c$ is a numerical constant.

  Because of the anomaly, the external gauge field cannot be quantized in the
above model. One way to cancel the anomaly
is to add new left-handed fermions with different U(1)
charges such that Tr$\,Q^3=0$. Note that in this case $F\Tilde{F}(x)$ remains
a non-trivial cocycle, but we have arranged its coefficient to be zero.

  We can also try a different approach, namely, we can attempt to make the
operator $F\Tilde{F}(x)$ cohomologically trivial by adding new external
fields with suitable chiral transformation properties.
Since $F\Tilde{F}(x)$ itself
is chirally invariant, the only way to obtain it as the chiral variation of
another operator is by adding a real scalar field $\f(x)$ transforming as
$$ \d_{\s} \f(x)=v \s(x) \,,\eqno(2.10)$$
where the constant $v$ has mass dimension one.

  Adding the new external field changes the cohomological problem. The new
BRST operator is
$$ \S' = \S + \Int{^4x} v\s(x) {\d\over\d \f(x)} \,.\eqno(2.11)$$
We can now restore the gauge invariance of the effective action by adding
a Wess-Zumino term
$$ \G'(A_{\m},\f) = \G(A_{\m}) - {c g^2\over v}\Int{^4x} \f(x) F\Tilde{F}(x)
\,,\eqno(2.12)$$
with
$$ \S'\, \G' = 0 \,.\eqno(2.13)$$

  In the context of ordinary gauge theories, adding the Wess-Zumino term
leads to a non-normalizable theory. (The Wess-Zumino term is nevertheless
useful in constructing low energy effective chiral lagrangians).
In gravity theories, we are faced from the beginning
with a non-renormalizable theory. The question of quantum consistency
can however be discussed independently\footnote{
  The non-renormalizability
problem can be solved if (super)gravity is the low energy limit of a more
fundamental theory.}.
  If adding a Wess-Zumino term makes the difference between a consistent
and an inconsistent theory we should opt for introducing it as a counter-term.

  Having obtained a gauge invariant effective action we now want to quantize
the external fields. Notice that because of its transformation law eq.~(2.10),
$\f(x)$ can be eliminated algebraically using the local chiral
invariance and hence it is a compensator. Consequently, there should exist an
alternative description of the final theory which does not contain $\f(x)$
and which has no local chiral invariance. At the moment we want
to know whether in adding the Wess-Zumino term we have guaranteed
the consistency of the final theory.

  In order to see what might go wrong let us first quantize only $\f(x)$.
To this end we functionally integrate
over $\f(x)$ with a suitable measure
\beqr{2.14}
  Z(A_{\m}) & = & \int \cd\f\, \exp \{i \G'(A_{\m},\f)+S_{\f}\} \NON
             & = & \int \cd\f \cd\Bar{\j} \cd\j\, \exp \{i S'_m\}
\,,\eeqr
where
$$ S'_m = S_{\j} + S_{\f} -
               {c g^2\over v} \Int{^4x} \f(x) F\Tilde{F}(x) \,,\eqno(2.15)$$
is the total matter action.
  The action $S_{\f}$ must be invariant under the local chiral
transformations eqs.~(2.7) and~(2.10). A suitable kinetic energy term is
$$ S_{\f} = \half \Int{^4x} (\pa{\m}\f - m A_{\m})^2  \,,\eqno(2.16)$$
where $m = gv$. The new Noether current obtained from the $A_{\m}$-variation
of the matter action is
$$ j'_{\m} = j_{\m} + v(\pa{\m}\f - m A_{\m})
           - {4c g^2\over v} \partial^{\n} (\Tilde{F}_{\m\n} \f)
\,.\eqno(2.17)$$
Notice that the $A_{\m}$-variation of the Wess-Zumino term
gives rise to an identically conserved current. Using
$$ \partial^{\m} j_{\m} = c g^2 F\Tilde{F} \,,\eqno(2.18)$$
as well as the $\f(x)$ field equation
$$ \bo \f = m\partial^{\m} A_{\m} - {c g^2\over v} F\Tilde{F} \,,\eqno(2.19)$$
we find that $j'_{\m}$ is conserved.

  A dispersive analysis of the triangle graph~[6] shows that the anomaly is
related
to a peculiar intermediate state containing two parallel fermions which
produce a $\d(q^2)$ discontinuity. Adding $\f(x)$ allows us to cancel the
anomaly because it produces a similar discontinuity but with an adjustable
coefficient.

  We can now quantize $A_{\m}$ in the usual way. The generating
functional of the complete quantum theory is
$$ Z = \int \cd A_{\m} \cd\f \cd\Bar{\j} \cd\j\, \exp \{i S'\}
\,,\eqno(2.20)$$
where
$$ S' = -{1\over 4}F^2 + S'_m  \,.\eqno(2.21)$$

  As anticipated, we can reformulate the resulting theory without $\f(x)$.
To this end, we use the local chiral invariance which is now maintained
also at the quantum level to set $\f(x) = 0$.
Fixing the local invariance is done as usual using the
Faddeev-Popov method. One multiplies the integrand of eq.~(2.20) by unity in
the
form of
$$ 1 = \int \cd\o\, J\, \d\left(\cf^{(\o)}(x)\right)\,.\eqno(2.22)$$
The notation $\cf^{(\o)}(x)$ stands for the gauge condition $\cf(x)$ rotated
by a local chiral transformation with parameter $\o(x)$.
As a gauge condition we choose $\cf(x)=\f(x)$. Recall that
$$ \f^{(\o)}(x)=\f(x)+v\o(x) \,.\eqno(2.23)$$
The jacobian is $J=1$. In choosing
this gauge condition we are using the fact that $\f(x)$  can be
identified with the parameter of the local gauge group. We now make a change
of variables which is a gauge transformation with parameter
$-\o(x)$. This results in a new
expression for the generating functional
\beqr{2.24}
 Z & = & \int \cd\o \cd A_{\m} \cd\f \cd\Bar{\j} \cd\j\,
      \d\left(\f(x)\right) \exp \{i S'\} \NON
   & = & \int \cd A_{\m} \cd\Bar{\j} \cd\j\, \exp \{i S\}\,,
\eeqr
where
$$ S = \Int{^4x} \left( -{1\over 4} F^2 + \half m^2 A^2
        + S_{\j}  \right)
\,.\eqno(2.25)$$
The action $S$ is obtained from $S'$ of eq.~(2.21) simply by setting
$\f(x) = 0$. In going from the first to the second line of eq.~(2.24) we have
dropped $\int\cd\o$ which is now an overall normalization constant.

  We recognize eq.~(2.25) as the Stueckelberg model~[1,2].
Notice that $A_{\m}$ has become a massive vector field.
In fact, one can identify $\f(x)$ with its longitudinal component.
The Wess-Zumino term has disappeared upon
setting $\f(x) = 0$, and the U(1) current that follows from the action $S$
is the original one $j_{\m}$. Thus, the current is anomalous, but this is now
harmless because the chiral symmetry is a global one.

  Now, suppose that the only thing one had to worry about is keeping the
invariance of the generating functional under local chiral
transformations. (As we will immediately see, this is not true if one wants
to obtain a meaningful quantum field theory). We could then define another
generating functional by not adding any kinetic term for $\f(x)$ at all!
Setting $S_{\f}=0$ in eq.~(2.15) we obtain
\beqr{2.26}
  \Hat{Z}(A_{\m}) & = & \int \cd\f\, \exp \{i \G'(A_{\m},\f) \} \NON
             & = & \int \cd\f \cd\Bar{\j} \cd\j\, \exp \{i \Hat{S}'_m\}
\,,\eeqr
where
$$ \Hat{S}'_m = S_{\j} -
               {c g^2\over v} \Int{^4x} \f(x) F\Tilde{F}(x) \,.\eqno(2.27)$$

  A glance at eq.~(2.27) reveals what goes wrong. $\f(x)$  occurs in
the action $\Hat{S}'_m$ as a Lagrange multiplier. Functionally
integrating over it gives rise to
$$ \Hat{Z}(A_{\m}) = \int \cd\Bar{\j} \cd\j\, \d\left(F\Tilde{F}(x)\right)
                       \exp \{i S_{\j}\} \,.\eqno(2.28)$$
Equivalently, the current that follows from the
action $\Hat{S}'_m$ is
$$ \Hat{j}'_{\m} = j_{\m} - {4c g^2\over v}\partial^{\n} (\Tilde{F}_{\m\n}\f)
\,.\eqno(2.29)$$
It satisfies
$$ \partial^{\m} \Hat{j}'_{\m} = c g^2 F\Tilde{F} \,,\eqno(2.30)$$
but the \rhs of eq.~(2.30) is set to zero by the $\f(x)$ field equation which
is
$F\Tilde{F}(x)=0$.

  Obviously, the constraint $F\Tilde{F}(x)=0$ is physically unacceptable.
Since $F\Tilde{F}(x)$ has non-vanishing matrix elements between the vacuum and
two photon states, enforcing this constraint would require the absence of two
photon states from the Hilbert space. But this is inconsistent with the
canonical commutation relations for $A_{\m}$ and hence $A_{\m}$
cannot be quantized. In fact, one cannot even require that the $A_{\m}$ field
equation should hold for $A_{\m}$ as a classical field,
because on a general $n$-fermion state one has  $F_{cl}\Tilde{F}_{cl}(x)\ne 0$.
Thus, requiring $F\Tilde{F}(x)=0$ plus the classical $A_{\m}$ field equation
gives rise to a trivial Hilbert space containing only the vacuum state.

  Suppose that, ignoring the unacceptable consequences of the action
$\Hat{S}'_m$ we proceeded as in the previous case. We could define
a generating functional
$$ \Hat{Z} = \int \cd A_{\m} \cd\f \cd\Bar{\j} \cd\j\, \exp \{i \Hat{S}'\}
\,,\eqno(2.31)$$
where
$$ \Hat{S}' = -{1\over 4}F^2 + \Hat{S}'_m  \,.\eqno(2.32)$$
Just like $Z$ of eq.~(2.20), the generating functional $\Hat{Z}$ is invariant
under local chiral transformation. We can therefore repeat the Faddeev-Popov
procedure to eliminate $\f(x)$. The result is
$$ \Hat{Z} = \int \cd A_{\m} \cd\Bar{\j} \cd\j\, \exp \{i \Hat{S}\}
\,,\eqno(2.33)$$
where
$$ \Hat{S} = \Int{^4x} \left( -{1\over 4} F^2
        + S_{\j}  \right) \,.\eqno(2.34)$$
At first glance it seems that fixing the local chiral invariance we have
obtained
a generating functional which is still invariant under local chiral
transformations\ldots But, of course, this is not true because
the local chiral invariance of $\Hat{S}$
is lost at the quantum level, which is precisely the anomaly we have tried
to avoid!

\vspace{5ex}
\noindent {\bf 3.~~Derivation of the on-shell constraint}
\vspace{3ex}

  The example of the previous section shows that in a theory which is
formulated using compensating fields, trivial cohomology is not a sufficient
condition for quantization. This result is not surprising. The
ability to consistently quantize a theory cannot depend entirely on its
algebraic symmetry structure.

  In addition to algebraic properties, what enters the diagrammatic analysis
of anomalies is the kinematics of the quantized fields. When all fields have
the kinetic energy terms appropriate to their spin,
the analysis of anomalies can be
reduced to an a algebraic problem. In the above example, the anomaly could
be eliminated either by adding more fermions with suitable
charges or by adding a propagating compensator. These two choices have
different
physical content but they are both consistent. What is common to both is that
the new fields change the matrix element of the current between {\it physical
states}. On the other hand, in the case of the non-propagating compensator no
change
occurred in this matrix element, and the anomaly was ``avoided'' by the
inconsistent requirement that the relevant two photon states be absent from
the Hilbert space.

  Since trivial cohomology is not a sufficient condition for quantization
in a theory formulated with compensators, the need arises for additional
criteria that will allow us to determine which theories can be consistently
quantized. In this section we will show that each compensating field gives rise
to a constraint. The constraint requires that the inhomogeneous term in the
field equation of the compensator vanish on the physical Hilbert space.
Thus, the content of these constraints determines
whether the theory can be quantized consistently. We will derive
the constraint for a compensator which can be eliminated
by fixing an abelian unphysical symmetry (henceforth denoted as abelian
compensator). We expect, however, that similar
results should hold in general.

  Before the actual derivation let us examine the content of the constraints
in the examples of the previous section. In the case of the
non-propagating compensator, the action contained only a term linear in
$\f(x)$.
The corresponding field equation was $F\Tilde{F}(x)=0$. This equation is
clearly a constraint even before one sets the compensator to zero and,
furthermore, it is obvious that this constraint
is physically unacceptable.

  The field equation for the propagating compensator is eq.~(2.19).
Setting $\f(x)=0$
in this equation we obtain the constraint
$$ m^2\partial^{\m} A_{\m} = c g^3 F\Tilde{F} \,.\eqno(3.1)$$
The easy way to verify that this constraint is physically sensible is to
examine the $A_{\m}$ field equation after the elimination of the compensator.
As follows from the action $S$ of eq.~(2.25), this field equation is
$$ \partial^{\n}F_{\n\m} + m^2 A_{\m} = g j_{\m} \,.\eqno(3.2)$$
Acting on this equation with $\partial^{\m}$ and using the anomaly
equation~(2.18) we obtain eq.~(3.1).

  The physics behind this constraint is that a
massive vector field contains only three degrees of freedom, but in order to
describe it in a Lorentz covariant way one needs the $A_{\m}$ field which
has four components. Thus, one of the components of $A_{\m}$ should
not be independent. If the massive vector field were coupled to a conserved
current its field equation would imply the vanishing of $\partial^{\m} A_{\m}$.
In general, however, the source current need not be conserved and the
$A_{\m}$ field equation allows us to express $\partial^{\m} A_{\m}$ in terms
of the divergence of the source current.

  We will now prove the existence of the on-shell constraint for an abelian
compensator. As we will see, the constraint is a consequence of the
invariance of the generating functional under the  corresponding
unphysical symmetry. It can be expressed in terms of the field equation
of the compensator because of its peculiar transformation law. (One might worry
whether the unphysical symmetry could be anomalous. As the example of the
previous section shows, any non-invariance
of the effective action can be cancelled by a counter-term which depends
linearly on the compensator. This property is obvious for an abelian
compensator and we expect it to be true in general).

  Consider a theory which is formulated using an abelian compensator $\f(x)$
as well as other fields. We denote all the other fields collectively as
$\c(x)$.
The generating functional
$$ Z = \int \cd\f \cd\c\, \exp \{i S(\f,\c)\}  \,,\eqno(3.3)$$
is invariant under local U(1) transformations with parameter $\o(x)$. We
assume that the $\f(x)$ transformation law is given by eq.~(2.23).
Let us separate the $\cd\f$ integration in eq.~(3.3) by rewriting it as
$$ Z = \int \cd\f\, Z(\f)\,, \eqno(3.4)$$
$$ Z(\f)=\int \cd\c\, \exp \{i S(\f,\c)\}  \,.\eqno(3.5)$$
$Z(\f)$ can be thought of as the generating functional where $\f(x)$ is an
external field. Notice, however, that it contains the complete action including
in particular possible kinetic terms for $\f(x)$.

  We observe that $Z(\f)$ too is invariant under local chiral transformations.
Because of the special transformation law of $\f(x)$,
the jacobian arising from the following change of variables in $Z(\f)$
$$ \c \to \c^{(\o)}  \,,\eqno(3.6)$$
is always the same as the jacobian arising from the
simultaneous change of variables in $Z$
$$ \left(\c,\f\right) \to \left(\c^{(\o)},\f^{(\o)}\right)
\,.\eqno(3.7)$$
By assumption, any non-trivial jacobian that may arise from the above change
of variables in $Z$ is cancelled by the non-invariance of the action. The
same statement is therefore true for $Z(\f)$. As a result
$$ Z(\f^{(\o)})=Z(\f) \,.\eqno(3.8)$$
But for $\f(x)$ a gauge transformation is simply a shift! Consequently,
$$ Z(\f + \d\f)=Z(\f^{(v^{-1}\d\f)})=Z(\f) \,.\eqno(3.9)$$
Thus, the invariance of $Z(\f)$ under local gauge transformations implies that
it is $\f$-independent. We rewrite this statement as
$$ { \d \over \d\f} Z(\f) = 0 \,. \eqno(3.10)$$

  We make contact with the previous section by noting that the result of
performing the Faddeev-Popov procedure can be written as
$$ \int \cd\f\, Z(\f) = \left( \int \cd\f\right) Z(0) \,.\eqno(3.11)$$
We see that equality in eq.~(3.11) holds in fact already at the level of the
integrand.

  Eq.~(3.10) is identical to the $\f(x)$ field equation obtained from
the generating functional $Z$ provided we treat $\f(x)$ as a classical field.
In particular, evaluating the \lhs of eq.~(3.10) at $\f(x)=0$ we obtain
the promised constraint. The above analysis can be repeated without change if
gauge invariant sources are added to the generating functional. As a result,
the constraint applies to the entire physical Hilbert space.

  Although we have derived the constraint for a single abelian compensator,
we expect that the same result should hold in general. In more complicated
situation (such as the superspace formulation of supergravity theories)
one may face a large number of compensators transforming in a complicated way
under a large number of unphysical symmetries. Nevertheless, the requirement
that all compensators can be eliminated algebraically implies a one-to-one
correspondence between the unphysical symmetries and the compensators.
To each compensator it should be possible to associate an unphysical
symmetry under which it transforms by a shift. Thus,
the above analysis can be repeated with the conclusion that each compensator
gives rise to the corresponding on-shell constraint.

\vspace{5ex}
\noindent {\bf 4.~~New minimal supergravity}
\vspace{3ex}

 As discussed above, in a theory formulated with
compensators, the invariance under the unphysical symmetries implies certain
constraints on the physical Hilbert space. The successful quantization of the
theory depends on one's ability to impose these constraint on the
operators of the theory without creating physically unacceptable consequences.

  In this section we apply the above general consideration
to {\it new minimal} supergravity. The simplest way to deal with the new
constraints is to avoid them altogether by formulating the theory without
compensators! For supergravity theories, this means that one has to use the
components formulation. As discussed below, new minimal supergravity has
the peculiar property that one of its auxiliary fields enters the action as
a Lagrange multiplier. Thus, even in the components formulation one field
equation is a constraint. We will show that, in general, both the invariance
under local SUSY transformations and the constraint are violated at the
quantum level.

 The physical fields of the supergravity multiplet are the tetrad
$e_{\m}^a$ and the gravitino $\j_{\m}^{\a}$.
On-shell, each of these fields have two degrees of freedom.
Off-shell, the tetrad has six degrees of freedom whereas the gravitino has
twelve. (One counts the number of components minus the number of local
symmetries which are gauged by the field).
A minimal set of auxiliary fields should therefore contain
six bosonic degrees of freedom.

  In global SUSY, \ie for chiral and vector multiplets, the auxiliary fields
are uniquely determined. In supergravity, on the other hand, there exist two
minimal sets of auxiliary fields. In ``old minimal'' supergravity~[5]
one introduces an axial vector $A_{\m}$,
a real scalar $S$ and a pseudoscalar $P$.

  In ``new minimal'' supergravity~[3] one introduces an axial {\it gauge} field
$A_{\m}$ and an antisymmetric tensor $a_{\m\n}$.
The $A_{\m}$ field gauges a global R-symmetry
of the on-shell formulation. Because of the local chiral invariance
it contributes $4-1=3$ degrees of
freedom off-shell. There is an additional gauge symmetry which acts only on
the antisymmetric tensor. Its transformation law is
$\d a_{\m\n} = \partial_{[\m}\L_{\n]}$.
Thus, $a_{\m\n}$ also contributes $6-3=3$
degrees of freedom. For the supergravity transformation rules of each
formulation we refer the reader to the original literature.

  In each off-shell formulation one can describes a
different set of on-shell theories. What
characterizes the on-shell models obtained from new minimal supergravity
is the presence of a global R-invariance. This restriction is absent in
old minimal supergravity which describes a much bigger class of models.

  Many on-shell theories can however be described in both off-shell
formulations. In particular, these include supergravity coupled to (abelian
or non-abelian) vector multiplets. We will show below
that {\it new minimal} supergravity coupled to vector
multiplets is anomalous. (In the presence of scalar multiplets anomaly
cancellation is possible). It is a commonly held view that,
in the absence of gauge anomalies, {\it old minimal} supergravity coupled to
matter gives rise to a consistent quantum theory. In our opinion,
a satisfactory proof of the consistency of old minimal supergravity
has not been given to date, and we prefer to leave it as an open question.
In any event, before we prove the existence of an anomaly
in new minimal supergravity there is an
apparent paradox which must be resolved. Using the field equations of the
auxiliary field one can superficially reduce both off-shell formulations to
the same on-shell theory. How can one off-shell formulations be inconsistent
without implying that the other one is inconsistent too?

  We claim that the two off-shell formulations should be regarded as two
{\it different} theories. The coupling of a quantum matter
system to external linearized supergravity requires the existence of
a conserved, symmetric energy-momentum tensor and a conserved spinor current.
Furthermore, these currents must belong to a supermultiplet which
forms an irreducible representation of global SUSY~[7,8].
The other currents in this multiplet
are sources for the auxiliary fields of the supergravity multiplet.
In particular, in both old and new minimal supergravity the field $A_{\m}$
couples to an axial R-current.

  Now, because of their different sets of auxiliary fields and
different supergravity transformations rules, the two off-shell formulations
define two different current algebras. The main difference is that the
new minimal current algebra~[3,10] requires the conservation of the
axial current whereas the old minimal current algebra~[7] does not.
As a result, the Ward identities of the two formulations
have different content. In the limit of weak supergravity fields they lead
to different predictions for the matrix elements of the currents between
physical states. Since these are physical observables,
we necessarily reach the conclusion that the two minimal formulations
are in fact different theories.

  How are these two options reflected in the on-shell formulation? At the
classical level both current algebras reduce to the superconformal current
algebra. The differences between them arise only at the quantum level and
reflect the existence of chiral and (super)conformal anomalies.
One effectively chooses between the two off-shell formulations by postulating
additional, {\it anomalous} terms in the transformation law of the
spinor and axial currents.

  The occurrence of anomalous terms in the transformation law of certain
composite operators in known as the Konishi anomaly~[9].
These terms are necessary
in order to maintain the supermultiplet structure at the quantum level~[9,10].
That the transformation law of the SUSY current itself must be modified
at the quantum level has not been fully appreciated in the literature.
(The transformation law of the axial current has been recently discussed in
ref.~[11]). This issue is discussed in more detail in a separate
publication~[12]. The fact that
new minimal supergravity is in general anomalous implies that, if one
postulates
the additional quantum terms appropriate to the current algebra of new minimal
supergravity, then one is bound to violate some on-shell Ward identities.
We consider as an open question whether all Ward identities are satisfied
if one postulates the additional quantum
terms appropriate to old minimal supergravity.

  Having clarified the differences between old and new minimal supergravity
we now proceed to show that new minimal supergravity is in general anomalous.
We will show that in new minimal supergravity,
a local supersymmetry anomaly is an immediate
consequence of the chiral anomaly. Thus, a necessary condition for the
consistency of new minimal supergravity is the existence of a conserved
R-current.

  Although the components formulation contains no compensating
fields, it turns out that the axial gauge field $A_{\m}$  enters the action
as a Lagrange multiplier. Its field equation is therefore a constraint.
Explicitly, it requires that the axial R-current be equal on-shell to the field
strength of the antisymmetric tensor $\e^{\m\n\s\t}\partial_\n a_{\s\t}$.
Since the latter is an identically conserved current, the constraint requires
that the axial current be conserved. Thus, the $A_{\m}$ field equation
-- the constraint -- is violated whenever the axial current is not conserved.
As a result, one can no longer use the field equations of the auxiliary fields
to reduce the off-shell formulation to the on-shell one.

  In addition to the new minimal supergravity multiplet let us now introduce a
vector multiplet $(B_{\m},\l,D)$. For a Yang-Mills multiplet there is an
additional adjoint representation index which we suppress.
The local new minimal SUSY algebra is~[3]
\beqr{4.1}
  \d_S(\h)\d_S(\z) \!\!\! & - & \!\!\! \d_S(\z)\d_S(\h) \, = \,
  \d_{gc}(-\x^{\m})+\d_S( {\k\over 2} \x^{\m}\j_{\m}) \NON & &
  + \, \d_L(-\x^{\m}\o_{\m ab}- {\k\over 2} \x^{c}V^d\e_{abcd})
  +\d_R(- {\k\over 2} \x^{\m}A_{\m})
  +\d_G(- g \x^{\m}B_{\m})
\eeqr
where
$$ \x^{\m} = 2i\, \Bar{\z}\g^{\m}\h  \,,\eqno(4.2)$$
$$ \o_{\m ab} = \o_{\m ab}(e) - {i \k^2\over 4}
   \left( \Bar{\j}_{\m} \g_a \j_b
        + \Bar{\j}_a \g_{\m} \j_b
        - \Bar{\j}_{\m} \g_b \j_a   \right)
\,,\eqno(4.3)$$
is the spin connection and
$$ V^{\m} = e^{-1} \e^{\m\n\s\t} \left( \partial_\n a_{\s\t} +
            {i \k\over 2} \Bar{\j}_{\n} \g_\s \j_\t \right)
\,.\eqno(4.4)$$
In eq.~(4.1), $\d_S(\h)$ is a local SUSY transformation where the
anti-commuting parameter $\h$ is a Majorana spinor. $\d_{gc}(-\x^{\m})$
is a general coordinate transformation with (commuting) parameter $\x^{\m}$.
The other terms on the \rhs of eq.~(4.1) stand for local  SUSY,
Lorentz, axial and gauge transformations with the specified field dependent
parameters. Notice that the role of these terms is to covariantize the local
translation with respect to all other local symmetries.
(In section~2 we introduced the BRST operator which is defined in terms of
ghost parameters with inverse statistics. Here we find it more convenient to
introduce parameters with normal statistics).
We also give the commutator of local SUSY and chiral transformations
$$ \d_S(\h)\d_R(\o)-\d_R(\o)\d_S(\h) = \d_S(i\o\g_5\h)  \,.\eqno(4.5)$$

  We now consider the quantization of the matter sector with the fields of
the supergravity multiplet as external sources. We intend to examine the
behaviour of the effective action
$$ \G = \G^{(0)} + \hbar \G^{(1)} + \ldots  \,,\eqno(4.6)$$
under local chiral and SUSY transformations.
A technical difficulty is that the gauge fixing terms needed for the
quantization of the gauge field $B_{\m}$ break SUSY explicitly.
However, this breaking disappears if one evaluates the matrix elements
of gauge invariant operators between physical states~[13].

  The tree level effective action respects
all classical symmetries. At the one loop level we may encounter
anomalies. We define the chiral and SUSY anomalies respectively by
$$ A_R(\o) = \d_R(\o) \G^{(1)}  \,,\eqno(4.7)$$
$$ A_S(\h) = \d_S(\h) \G^{(1)}  \,,\eqno(4.8)$$
with the understanding that a projection onto physical states is applied
(\eg by using the LSZ reduction formalism) in eq.~(4.8).

  In four dimension there are no gravitational and Lorentz anomalies.
We furthermore assume the absence of gauge anomalies. Thus,
$$ \d_{gc}\G=\d_{L}\G=\d_{G}\G=0  \,.\eqno(4.9)$$
We now obtain the Wess-Zumino consistency conditions by applying the
operatorial
equations~(4.1) and~(4.5) to $\G^{(1)}$. We obtain, respectively
$$ \d_S(\h)A_S(\z)-\d_S(\z)A_S(\h) =
   A_S( {\k\over 2} \x^{\m}\j_{\m})+A_R(- {\k\over 2} \x^{\m}A_{\m})
\,,\eqno(4.10)$$
$$ \d_S(\h)A_R(\o)-\d_R(\o)A_S(\h) = A_S(i\o\g_5\h)  \,.\eqno(4.11)$$

  Eq.~(4.10) implies that the SUSY anomaly cannot vanish usless the chiral
anomaly does. Thus, new minimal supergravity cannot be quantized
if the R-current is anomalous, for in this case the SUSY current is anomalous
too.

  The Wess-Zumino consistency condition~(4.11) can be easily solved.
We write
$$ A_S = A_S^{cov} + A_S^{non} \,.\eqno(4.12)$$
The two terms on the \rhs of this equation denote the covariant
and non-covariant parts of the SUSY anomaly with respect to
axial transformations. We now demand
$$ \d_S(\h)A_R(\o)-\d_R(\o)A_S^{non}(\h) = A_S^{non}(i\o\g_5\h)
\,,\eqno(4.13)$$
$$ -\d_R(\o)A_S^{cov}(\h) = A_S^{cov}(i\o\g_5\h)  \,.\eqno(4.14)$$
An explicit solution for $A_S^{non}$ is given below. The only restriction on
$A_S^{cov}$ that follows from eq.~(4.14) is that its chiral charge is minus
one. We have not been able so far to find an explicit expression for
$A_S^{cov}$.

  Before we proceed to the superspace formulation let us examine the
possibility of anomaly cancellation. The axial charge of the gravitino
$\j_{\m}$ and the gaugino $\l$ is one. $A_{\m}$ transforms as a gauge field
whereas all other fields of the supergravity and vector multiplets are
chirally inert. If scalar multiplets are added, the chiral charges
of the scalar, fermion and auxiliary components are $q$, $q-1$ and $q-2$
respectively, where $q$ is the chiral weight of the
multiplet and is a-priori a free parameter.

  The R-current anomaly consists of three independent contributions
$F\Tilde{F}$, $R\Tilde{R}$ and $G\Tilde{G}$. $F_{\m\n}$ and $G_{\m\n}$ are
respectively the field strength of $A_{\m}$ and $B_{\m}$. $R_{\m\n}^{ab}$ is
the Riemann tensor.
The field strength of the auxiliary field $A_{\m}$ should vanish on-shell~[5],
and so the term $F\Tilde{F}$ is most likely harmless.
This leaves two independent anomalies $G\Tilde{G}$ and $R\Tilde{R}$
to worry about. In particular, we conclude that new-minimal supergravity
coupled to a Yang-Mills multiplet is always anomalous. Upon adding
scalar supermultiplets, anomaly cancellation becomes possible.
If the additional scalar multiplets have only kinetic terms their chiral
weights remain as free parameters.
(However, different assignments of chiral weights give rise to different
lagrangians~[3]). Thus, for instance, if one couples supersymmetric QCD with
two
quark flavours to new minimal supergravity, the chiral weights of the two
quark families can be adjusted to achieve anomaly
cancellation. However, if one adds an (R-covariant) superpotential the chiral
weights of the scalar multiplets are fixed and anomaly cancellation becomes
more involved.

 Explicit solutions for the non-covariant part of the SUSY anomaly
can easily be found. The non-covariant part that correspond
to $G\Tilde{G}$ is
$$ A_S^{non} = -2i\k \Int{^4x} A_{\m} \Tilde{G}^{\m\n} \Bar{\h}\g_{\n}\l
\,.\eqno(4.15)$$
Analogous expression exists for the gravitational term with $\l$ replaced
by the SUSY variation of the spin connection.
The anomaly~(4.15) implies that SUSY is violated in the correlator
of the SUSY current with an axial current and the source currents for the
gluon and the gaugino. SUSY violation in this Ward identity is
in fact a consequence of the ``wrong'' transformation laws postulated
for the SUSY and axial currents. One can check that there is no anomaly
in the corresponding Ward identity in old minimal supergravity.

\vspace{5ex}
\begin{flushleft}
{\bf 5.~~The relation between superspace formulation \\
\hspace{7mm}and the components formulation}
\end{flushleft}
\vspace{3ex}

  In the previous section we showed that local SUSY is anomalous in new minimal
supergravity whenever the chiral symmetry is anomalous. Thus, new minimal
supergravity cannot be quantized if the {\it axial current} is not conserved.

  The question arises whether one could do better in the superspace formulation
of new minimal supergravity. Answering this question is somewhat tricky because
in superspace the distinction between new and old minimal supergravity lies
in fine detail. Certain manipulations in superspace can effectively take us
from new to old minimal supergravity (or to a non-minimal theory).
With this reservation in mind our answer
to the above question is on the negative.

  The crucial observation that follows from our general discussion  on
compensators is the following. In field theory, one can always reformulate
a given theory using compensators. But when one introduces a new
compensating field into the action, {\it the generating functional for the
original fields remains independent of the new compensator}
(recall $Z(\f)$ of section~3). Thus, setting all compensating
fields to zero does not change the generating functional
(except for an unimportant normalization constant, see eq.~(3.11)).
Therefore, instead of asking what are the detailed manipulations one does in
superspace, we can ask directly what effect they could have on the components
formulation after the compensating fields have been eliminated.

  For the case at hand, the effect of possible superspace counter-terms
depends on whether the chiral compensator is propagating or not.
If the compensator is non-propagating, no trace of it
will be left in the Wess-Zumino gauge.
This means that the superspace theory is inconsistent, because its generating
functional is {\it equal} to the generating functional of the components
formulation which does not describe a consistent quantum theory.
If the chiral compensator is propagating, we end up with a {\it different}
lagrangian at the component level
which has no local chiral invariance. But this means
that the components formulation is now {\it old minimal} supergravity!
The modifications needed in order to obtain a propagating chiral compensator
have the effect of replacing new minimal by old minimal supergravity at the
superspace level.

  Gates, Grisaru and Siegel~[4] have discussed in detail the properties of the
various supergravity models in superspace. Our results concerning the
inconsistency of new minimal supergravity agree with their conclusions and we
will not repeat their analysis here.

  Few comments are however in place. Let us first recall what are
the basic objects needed to describe a supergravity theory in superspace.
One introduces the vielbein $E_M^A$ and the spin superconnection $\f_{MAB}$.
They are responsible for invariance under superdiffeomorphisms and
superlocal Lorentz transformation respectively.
(In this section we follow the notation of ref.~[14]).
In the Wess-Zumino gauge for supergravity one can identify some of their
lowest components with the physical fields.
Denoting $\th=\bar\th=0$ componets by $|$, the tetrad is $E_m^a|$,
the gravitino is $E_m^\a|$  and the spin connection is $\f_{mab}|$. Also,
$$ E_\m^a| = E_\dt\m^a| = E_\m^\dt\a| = E_\dt\m^\a| = 0 \,,\eqno(5.1)$$
$$ E_\m^\a| = \d_\m^\a \,,\eqno(5.2a)$$
$$ E_\dt\m^\dt\a| = \d_\dt\m^\dt\a \,.\eqno(5.2b)$$

  The vielbein and the spin superconnection are sufficient in order to describe
old minimal supergravity in superspace. In order to describe new minimal as
well as non-minimal models one
also introduce a U(1) superconnection $\G_M$ which gauges U(1)
rotations in the fermionic tangent space. Its generator is
$$ X_a^b = 0 \,,\eqno(5.3)$$
$$ X_\a^\b = i\d_\a^\b  \,,\eqno(5.4a)$$
$$ X_\dt\a^\dt\b = -i\d_\dt\a^\dt\b  \,.\eqno(5.4b)$$
$\G_m|$ is closely related to the chiral
gauge field of new minimal gravity, see below.

  It turns out that the superspace formulation has an additional local
invariance under so-called super-Weyl transformations. super-Weyl
transformations form an abelian group parametrized by a chiral superfield.
Their descendants in the component formulation are conformal
transformations.

  The above geometrical objects carry superspace indices and so they contain
a large number of superfields. These superfields are not all independent.
One relates them by postulating invariant superspace constraints\footnote{
  The reader should not mix the invariant superspace
constraints with the on-shell constraints of the previous
sections. The former are {\it postulated} in order to reduce the number of
independent superfields needed to describe the theory. The latter are {\it
derived} from the invariance under the unphysical symmetries of the superspace
formulation.}.
The introduction of invariant constraints is necessary, for instance, in order
to avoid the appearance of several independent connections for a single local
symmetry at the component level. Most of the  constraints
leave the invariance under Weyl transformations intact. One more
constraint is needed in order to reduce the theory from conformal
to Poincar\'{e} supergravity.
Depending on which constraint is chosen, one obtains old minimal,
new minimal or a non-minimal model already at the superspace level.

  Solving the Poincar\'{e} supergravity constraint gives rise to a smaller
local
symmetry group with a more complicated structure.  It is more
convenient to express this constraint in terms of an additional
superfield whose components are all compensators. The compensating superfield
is built out of the vielbein and the spin superconnection and there is no need
to introduce it as an additional independent object.
It compensates in a supersymmetric way for some or all of the Weyl
transformation, thus making them unphysical symmetries.

  Our first comment concerns the identification of the local chiral symmetry
of new minimal supergravity with one of the superspace symmetries.
To the single local R-symmetry that one has in the component formulation,
there correspond {\it four} local symmetries (or three if a reality
condition is imposed) in superspace!

  The most obvious descendants of the R-symmetry of global SUSY
is what we call R-superdiffeomorphisms. The parameters of these
superdiffeomorphisms are
$$  \x^m (x,\th, \bar{\th}) = 0  \,,\eqno(5.5)$$
$$  \x^{\m} (x,\th, \bar{\th}) = i \o(x) \th^{\m}  \,,\eqno(5.6a)$$
$$  \Bar\x^{\dt\m} (x,\th, \bar{\th}) = -i\Bar\o(x)\bar\th^{\dt\m}
\,.\eqno(5.6b)$$
The transformation law  under R-superdiffeomorphisms is
therefore determined by the {\it curved} fermionic indices. (Imposing a
reality condition amounts to setting $\Bar\o(x)=\o^*(x)$).

  Another chiral symmetry belongs to the super-Weyl group. It is
parametrized by the imaginary part of the lowest component of its parameter
superfield. Finally, there is the chiral symmetry gauged by
the ordinary connection which is contained in the U(1) superconnection.
(In old minimal supergravity the U(1) superconnection is not
an independent object).

  In order to reduce the number of local chiral symmetries to one
(or to zero in the case of old minimal supergravity), the superspace
formulation should contain several chiral compensators.
Two chiral compensators are easily found.
Following ref.~[1], we parametrize the part of the vielbein with only dotted
(only undotted) indices as
$$ E_{\m}^{\a} =  N_{\m}^{\a}\exp \{\F\} \,,\eqno(5.7a)$$
$$ E_{\dt\m}^{\dt\a} = \Tilde{N}_{\dt\m}^{\dt\a}\exp \{\Tilde\F\}
\,,\eqno(5.7b)$$
where $N_{\m}^{\a}$ and $\Tilde{N}_{\dt\m}^{\dt\a}$ are SL(2,C) supermatrices.
Notice that the superfields $\F$ and $\Tilde\F$ are not ordinary scalar
superfields. Their transformation law under superdiffeomorphisms is determined
implicitly by the transformation law of the vielbein.

  The chiral compensators are $Im\F|$ and $Im\Tilde\F|$.
(The real parts are compensators for scale transformations).
Indeed, in the Wess-Zumino gauge (see eq.~(5.2))
one has $\F|=\Tilde\F|=0$ (as well as $N|=\Tilde{N}|=I$).
Notice that the compensators are basically the logarithm of the
fermionic vielbein. Usually, the phase and
modulus of a given field are not  ``legal'' fields. Using them gives rise
to singularities due to the behaviour of the logarithm at zero. However,
the fermionic vielbein is expanded around the identity matrix (it is equal to
the identity matrix in the Wess-Zumino gauge) and so the parametrization~(5.7)
is legitimate.

  Since the superspace formulation contains several ``copies'' of the chiral
symmetry, the transformation law of a physical field such as the gaugino
under the superspace chiral symmetries is ambiguous. The only requirement is
that it transforms suitably under the linear combination which survives in
the components formulation. For instance, the definitions
$\l_{\a}=W_{\a}|$ and $\l_{\a}=\exp \{i Im \F\} W_{\a}|$ are both acceptable,
but they imply different transformation roles for the gaugino
under the superspace symmetries.

  With different definitions for the gaugino, the chiral anomaly will appear
as a breakdown of the invariance under different superspace symmetries. We can
then build Wess-Zumino terms using the chiral compensators.
Thanks to the one-to-one correspondence between compensators and the extra
symmetries of the superspace formulation,
we can always limit the quantum non-invariance
to a single chiral symmetry, say, to the chiral symmetry gauged
by the U(1) superconnection. An interesting question is whether one can
restore the invariance under {\it all} chiral symmetries at
the superspace level. At the moment we do not exclude this possibility.
Notice that our statement regarding the equality of the superspace and
components' generating functionals is not effected by this uncertainty,
because it depends only on the quantum invariance under the extra
symmetries which are eliminated in the Wess-Zumino gauge.

  Our second comment concerns the important issue of the consistency of
{\it old minimal} supergravity. A cohomological analysis of the
superspace formulation of old minimal supergravity was done by
Bonora, Pasti and Tonin~[15]. They were able to show that superdiffeomorphisms
and the super-Lorentz group are cohomologically trivial. This result
alone implies that among the superspace symmetries of old minimal supergravity
only the super-Weyl group could possibly be anomalous, and that its
anomaly can be brought into a supersymmetric form. Consequently,
using the compensating superfield of old minimal supergravity we can cancel the
super-Weyl anomaly without generating superdiffeomorphism or super-Lorentz
anomalies.
We therefore anticipate that the entire superspace cohomology of old minimal
supergravity is trivial.

  Because of the presence of compensators, the triviality of the superspace
cohomology does not guarantee the consistency of old minimal
supergravity. On the other hand, in the components formulation there are no
compensators, and so a proof that
the supergravity transformations of the components formulation are
cohomologically trivial would imply that old minimal supergravity is
consistent.
Unfortunately, this cohomological problem is considerably more complicated
and very little is known about it.

  The following scenario illustrates what could go wrong.
It is known from explicit calculations~[16]
to lowest order in the gravitational coupling
constant and with the graviton and the gravitino taken on-shell,
that the conformal anomalies arising from the quantization of
scalar or vector multiplets in an external supergravity background have
the supersymmetric form $W^{\a\b\g} W_{\a\b\g}$.
This still leaves open the possibility that contributions which vanish on-shell
to this order are not supersymmetric. (For instance, it has not been checked
whether contributions proportional to the \lhs of the gravitino field equation
occur with the right coefficient as required by the supermultiplet structure).
We point out that there is no conceptual difficulty in calculating these
off-shell contributions because  the supergravity fields are considered as
external sources. In the event that such terms do not have the desired
coefficients, the supermultiplet structure of the conformal anomalies would
be destroyed. This, in turn, would imply the existence of a non-trivial cocycle
in the components formulation of old minimal supergravity. It is likely that,
as a result, a local SUSY anomaly on-shell will appear at a
higher order in the gravitational coupling constant.

  Now, because of the presence of compensators,
the value of the conformal anomalies in the superspace formulation
is arbitrary. Indeed, we can use the conformal compensators to
build counter-terms which eliminate the conformal anomalies
altogether. Furthermore, we can do so regardless of whether or not
the conformal anomalies come with the right coefficients to
form a supermultiplet. In the event that the supermultiplet structure
is violated, it is not unreasonable to expect that
the above counter-terms will restore super-Weyl and
superdiffeomorphism invariance simultaneously. According to our general
discussion, the inconsistency would then show up as an inability to impose
the constraints generated by the component conformal compensators on the
physical Hilbert space.

  In trying to verify whether the above scenario has anything to do with
the true properties of supergravity theories, one should not make use of
manifestly supersymmetric methods such as dimensional
reduction. In fact, it is known that dimensional reduction is not a
consistent regularization scheme~[17]. It contains inherent ambiguities related
to objects such as $\e_{\m\n\s\t}$ and $\g_5$. Although in many cases
it gives the same results as dimensional continuation (up to finite
counter-terms), it is bound to fail precisely when there is an anomaly
behind the corner. For instance, if one were to calculate the triangle graph
using dimensional reduction, one would have found the wrong result
that the axial and vector currents can be conserved simultaneously.
Some ad-hoc modifications are needed in order to obtain the chiral anomaly
correctly, and there is no guarantee that dimensional reduction will not
fail in other circumstances as well.

  Other methods exist which partially solve the problem of
regularizing a theory while preserving global or local supersymmetry.
However, we are not aware of any regularization method which excludes the
above scenario. We point out that it is difficult to exclude this scenario even
for a scalar multiplet. While Pauli-villars regularization is available which
preserve global SUSY, it is not obvious how to generalize it to a locally
supersymmetric regularization for the component formulation.
The above questions are currently being investigated.

\vspace{5ex}
\centerline{\bf Acknowledgement}
\vspace{3ex}
I would like to thank L.~Bonora, S.~Elitzur, M.~Grisaru and A.~Schwimmer
for discussions.

\vspace{5ex}
\centerline{\rule{5cm}{.3mm}}

\newpage
\centerline{\bf References}
\vspace{3ex}
\newcounter{00001}
\begin{list}
{[~\arabic{00001}~]}{\usecounter{00001}
\labelwidth=1cm}

\item S.G.~Gates, M.T.~Grisaru, M.~Ro\br{c}ek and W.~Siegel, {\it Superspace},
Benjamin/Cummings Publ., 1983.

\item M.T.~Grisaru and B.~de Wit, in {\it Quantum Field Theory and Quantum
Statistics}, Vol.~2, ads. I.A.~Batalin {\it et al.} 1985.

\item M.~Sohnius and P.~West, \PLB{105} (1981) 353; \NPB{198} (1982)493.

\item S.G.~Gates, M.T.~Grisaru and W.~Siegel, \NPB{203} (1982) 189.

\item K.S.~Stelle and P.~West, \PLB{74} (1978) 330.
S.~Ferrara and P.~van Nieuwenhuizen, \PLB{74} (1978) 333.
P.~van Nieuwenhuizen, \PRP{68} (1981) 189.

\item Y.~Frishman, A.~Schwimmer, T.~Banks and S.~Yankielowicz,
\NPB{177} (1981) 157.

\item S.~Ferrara and B.~Zumino, \NPB{87} (1975) 207.

\item V.~Ogievetsky and E.~Sokatchev, \NPB{124} (1977) 309;
S.~Ferrara and B.~Zumino, \NPB{134} (1978) 301.

\item K.~Konishi, \PLB{135} (1984) 439.

\item T.E.~Klark, O.~Piguet and K.~Sibold, \NPB{143} (1978) 445; 159 (1979) 1.

\item H.~Itoyama B.~Razzaghe-Ashrafi, Stony Brook preprint ITP-SB-91-40.

\item Y.~Shamir, in preparation.

\item B.~de Wit and D.Z.~Freedman \PRD{12} (1975) 2286.

\item J.~Wess and J.~Bagger, {\it Supersymmetry and Supergravity},
Princeton Univ. press, 1983.

\item L.~Bonora, P.~Pasti and M.~Tonin,  \NPB{252} (1985) 458.

\item L.F.~Abbott, M.T.~Grisaru and H.J.~Schnitzer, \PLB{73} (1978) 71.

\item W.~Siegel, \PLB{94} (1980) 37.
G.~Bonneau, \PLB{96} (1980) 147.
L.V.~Avdeev, \PLB{117} (1982) 317.

\end{list}

\end{document}